\documentclass[conference]{IEEEtran}
\usepackage{graphicx,epsfig,epstopdf,amsmath,amssymb,enumerate} 
\usepackage{cite}  
\usepackage{multicol}

\begin{document}
\title{Intra-cluster Characteristics of 28 GHz Wireless Channel in Urban Micro Street Canyon}
\author{Shangbin~Wu, Sooyoung Hur, Kuyeon Whang, and Maziar Nekovee \\
\small{Samsung R\&D Institute UK, Communications House, Staines-upon-Thames, TW18 4QE, United Kingdom.} \\
Email: \{shangbin.wu, sooyoung.hur, kuyeon.whang, m.nekovee\}@samsung.com

}
\maketitle
\begin{abstract}
This paper investigates intra-cluster channel characteristics of non line-of-sight (NLOS) 28 GHz channels in street canyon scenarios. These channel characteristics include cluster numbers, number of subpaths within each cluster, intra-cluster delay spreads, and intra-cluster angular spreads. Both measurement and ray tracing results are presented and compared. Furthermore, distribution fittings are performed and models and parameters for different intra-cluster channel characteristics are proposed.

\end{abstract}

\IEEEpeerreviewmaketitle


\section{Introduction}
It is expected that the emerging fifth generation (5G) wireless access systems are able to provide unprecedented data rate for a large number of users. A wide range of high speed applications such as streaming high-definition videos will be supported by 5G. The aggregate data rate of 5G is required to be roughly a thousand times larger than that of the current fourth generation (4G) wireless systems \cite{whitepaper}--\hspace{-0.001cm}\cite{Gupta15}. This performance indicator is both appealing and challenging. 

Millimeter-wave (mmWave) communications have been considered as one of promising techniques for 5G because there are large unallocated spectral resources in mmWave frequency bands. Typical bandwidths in mmWave frequency bands are ranging from $0.5$ to $2$ GHz. By properly using these bandwidths, the targeted data rate in 5G can be achieved. For the purposes of mmWave communication system design, understanding of propagation characteristics is of crucial importance. It was stated in \cite{mmMAGIC_D2dot1} that volume scattering may be reduced in mmWave band due to the fact that penetration depth is inversely proportional with frequencies. Also, objects become large with respect to smaller wavelengths in mmWave bands so that they mainly contribute to specular components. Pathloss at $28$ GHz was studies in \cite{Sulyman14} and \cite{Rappaport13}, showing that pathloss at 28 GHz is severe and beamforming is required to compensate attenuations due to high frequency. Besides large-scale channel characteristics, small-scale channel characteristics at 28 GHz were discussed in \cite{Azar13} and \cite{Samimi13}. Multipath delay spreads for both line-of-sight (LOS) and non LOS (NLOS) environments at 28 GHz in \cite{Azar13} were in the order of hundreds of nanoseconds, which were smaller than those of channels at below 6 GHz. Angles of arrival (AoAs) and angles of departures (AoDs) were analyzed under the assumption of steerable beam antennas in \cite{Samimi13}. It can be observed that angular spreads were smaller due to the use of steerable beam antennas. Channel models based on the 3GPP three dimensional (3-D) spatial channel model (SCM) \cite{25996} at 28 GHz were developed in \cite{Hur15}. The channel parameters for this channel model were extracted via measurement data. Although extensive channel measurement results and channel characteristic analyses at 28 GHz have been reported in the literature, the investigation on intra-cluster channel characteristics at 28 GHz are missing. Intra-cluster characteristics are essential in mmWave channels, which is caused by high bandwidth and high time resolution in mmWave bands with large bandwidth. These characteristics include the number of subpaths within each cluster, intra-cluster delay spreads, and intra-cluster angular spread. In this paper, the abovementioned intra-cluster characteristics in a NLOS street canyon at 28 GHz are not only extracted via measurements but also via ray tracing simulations. The focus of this paper is in NLOS scenarios due to limited measurement results. LOS scenarios will be studied in future work.

The contributions of this paper are two-fold. First, intra-cluster characteristics of ray tracing simulations and measurements are compared. Their alignment demonstrates that ray tracing results can be used as reference when measurement results are not available. Second, statistics of intra-cluster characteristics are studied. Distributions are used to fit these characteristics, which can be employed in channel model development for above 6 GHz channels.
 
The remainder of this paper is organized as follows. Section~\ref{sec_Environment_settings} gives the general description of the ray tracing and measurement environment as well as their configurations. Section~\ref{sec_Intra_cluster_characteristics} presents definitions of intra-cluster characteristics. Both ray tracing simulation results and measurement results and their comparisons are discussed in Section \ref{sec_results_and_analysis}. Conclusions are drawn in Section~\ref{conclusion_section}.
\section{Environment settings} \label{sec_Environment_settings}
\subsection{Experiment environment description}
\begin{figure}[t]
\centering
\includegraphics[width=3.5in]{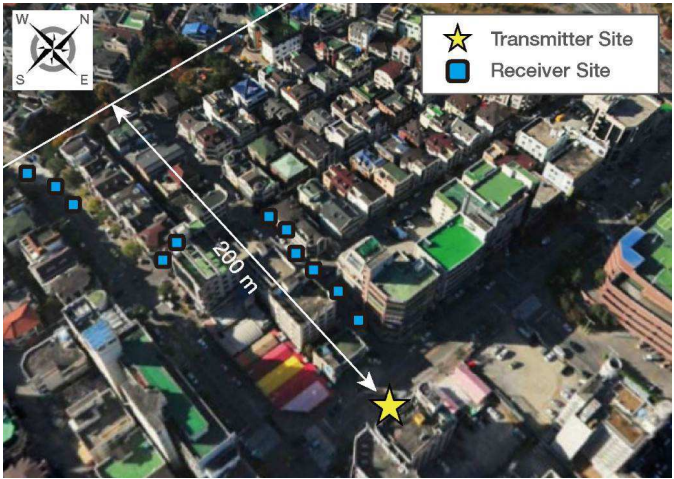}    
\caption{Bird's view of Daejeon (UMi street canyon) \cite{Hur15}.}
\label{fig_Measurement_locations}
\end{figure}

The experiment environment is in Daejeon, Korea. Streets and buildings are the main components in the environment. The distance between the transmitter and receiver is up to 200 m. The experiment environment is categorized as urban micro (UMi) street canyon. Several features of UMi street canyon scenarios have been identified in \cite{mmMAGIC_D2dot1}. UMi street canyon scenarios have high user density. Also, the users are mainly pedestrians or slow vehicular users. Buildings on both sides of the street have  normally four to seven storeys. The length of a street is in the order of 100 m. In addition, street furniture such as lampposts, traffic signs, and trees is typically seen in the environment. A bird's view of Daejeon is illustrated in Fig.~\ref{fig_Measurement_locations}.

\subsection{Ray tracing settings}
The ray tracing simulation was performed via the shoot-and-bounce method using the Wireless InSite \cite{RemCom} software. Rays with $0.1^\circ$ degree angular spacing were launched at the transmitter. Propagation effects on these rays such as reflections, diffractions, and penetrations were modeled using geometrical optics (GO) and uniform theory of diffraction (UTD). Then, the rays were traced up to a maximum number of reflections, diffractions, and penetrations until they reached the receiver. At the receiver side, a typical threshold 250 dB of propagation loss was assumed such that rays attenuated by more than 250 dB will not be considered in at the receiver. Delays, power, and angular information were the outputs of the ray tracing simulation.

\begin{figure}[t]
\centering
\includegraphics[width=3.5in]{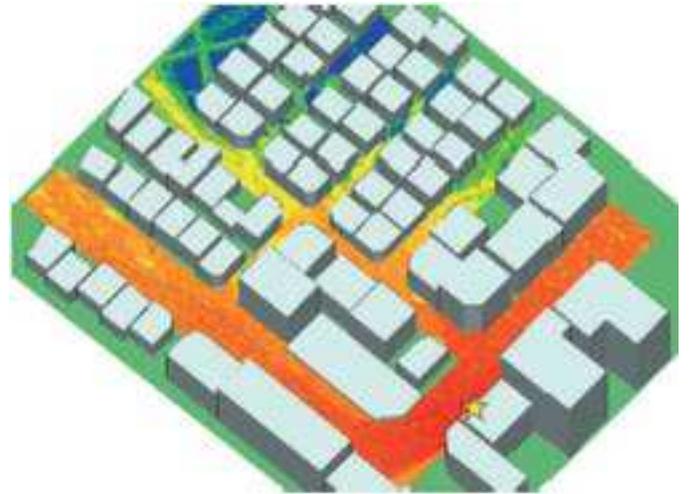}    
\caption{3-D geometrical layout of Daejeon for ray tracing simulations.}
\label{fig_DaejeonRaytracing}
\end{figure}

To mimic realistic environment as accurately as possible, 3-D geometrical information needs to be input to the ray tracing simulator. The 3-D layout of Daejeon is shown in Fig. \ref{fig_DaejeonRaytracing}. Moreover, frequency-dependent material properties were considered to model propagation characteristics at different frequencies. In this simulation, concrete was assumed for buildings and wet earth was assumed for the ground. In this case, permittivity values at 28 GHz for concrete and wet earth are $\epsilon=6.5$ and $\epsilon=15$, conductivity values at 28 GHz for concrete and wet earth are $\sigma=0.668$ and $\epsilon=1.336$ \cite{ITURP2040}. It can be argued that surfaces of buildings may contain other materials such as wood and glasses. However, the inclusion of other materials will introduce extra complexity. Also, it can be observed in later paragraphs that concrete only for buildings is able to present results well aligned with measurement.

Other input parameters such as maximum numbers of reflections, penetrations, diffractions, and rays in total were configured as Table \ref{tab_raytracing_setting}. Larger values can be used for better approximation of the realistic environment, at the cost of longer simulation time.

\begin{table}[ht]
\caption{Ray tracing simulation settings.}
\center
\normalsize
    \begin{tabular}{|c|c|}
    \hline
    Parameter  & Value  \\ \hline
        Ray spacing  & $0.1$ degree  \\ \hline
        Maximum number of reflections & 12  \\ \hline
                Maximum number of penetrations & 2 \\ \hline
                Maximum number of diffractions & 1  \\ \hline
                Maximum number of rays in total & 40   \\ \hline

    \end{tabular}
    \label{tab_raytracing_setting}
\end{table}

\subsection{Measurement settings}

In addition to ray tracing simulations, a measurement campaign was performed in the same environment. The channel was measured via a wideband radio channel sounder, which transmits 250 Mega chip-per-second (Mcps) pseudonoise (PN) sequences. 
The transmitter was located at a fixed position, which was $15$ m above ground. Receivers were placed in thirty eight different locations ($1.6$ m above ground) in street canyons, which are illustrated in Fig. \ref{fig_Measurement_locations}. 

Horn antennas with $24.5$ dBi gain and 10-degree half-power bandwidth were used in both the transmitter and receiver. These antennas can be rotated pointing various directions in azimuth and elevation automatically via a synchronized triggering signal. Both the transmitter and receiver scanned in the azimuth and elevation directions. The transmitter scanned from $-\pi$ to $\pi$ in the azimuth and from $-\frac{\pi}{3}$ to $\frac{\pi}{3}$ in the elevation. The receiver scanned from $-\frac{\pi}{3}$ to $\frac{\pi}{3}$ in the azimuth and from $-\frac{2\pi}{9}$ to $\frac{\pi}{18}$ in the elevation. Other specifications of the channel sounders are listed in Table \ref{tab_measurement_setting}. Measurement results were synthesized to generate omni-directional channel characteristics. In each measurement location, channel characteristics of rays such as power, delay, and angles were recorded.


\begin{table}[ht]
\caption{Channel measurement settings.}
\center
\normalsize
    \begin{tabular}{|c|c|}
    \hline
    Parameter  & Value  \\ \hline
        Carrier frequency  & $27.925$ GHz  \\ \hline
        Signal bandwidth & $250$ MHz  \\ \hline
                Transmit power & $29$ dBm \\ \hline
                Antenna type & Horn antenna  \\ \hline
                Antenna Gain & 24.5 dBi   \\ \hline
                 Antenna beamwidth & 10 degrees \\ \hline
                                 No. receiver locations & 38 \\ \hline

    \end{tabular}
    \label{tab_measurement_setting}
\end{table}
\section{Intra-cluster characteristics} \label{sec_Intra_cluster_characteristics}

In a 3-D propagation environment, a proper channel model should consider accurate space-time characteristics and impacts from angles in both azimuth and elevation. A generic structure for this channel model was presented in \cite{Maltsev10}. The channel impulse response can be expressed as
\begin{align}\label{received}
h(t,\mathbf{\Phi},\mathbf{\Theta})=\sum\limits_{n=1}^N \sum\limits_{k=1}^{M_n}a_{nk}\delta(t-\tau_{nk})\delta(\mathbf{\Phi}-\mathbf{\Phi}_{nk})\delta(\mathbf{\Theta}-\mathbf{\Theta}_{nk}).
\end{align}  
Subscripts $({\cdot})_{nk}$ are representing the $k$th subpath within the $n$th cluster. In addition, $a$ and $\tau$ are the complex gain and delay of a subpath. Vectors $\mathbf{\Phi}_{nk},\mathbf{\Theta}_{nk}$ are AoD and AoA directions, respectively. The AoD direction $\mathbf{\Phi}_{nk}=[\phi^{\mathrm{A}}_{nk}\: \phi^{\mathrm{E}}_{nk}]$ is a vector consisting of azimuth and elevation AoDs. Similarly, the AoA direction $\mathbf{\Theta}_{nk}=[\theta^{\mathrm{A}}_{nk}\: \theta^{\mathrm{E}}_{nk}]$ is a vector consisting of azimuth and elevation AoAs. It should be noticed that different numbers of subpaths are assumed in different clusters. Therefore, the number of subpaths $M_n$ within the $n$th cluster is depending on $n$. The number of clusters is denoted as $N$. 

\subsection{Number of clusters}
After obtaining data of subpaths via ray tracing or measurement, clustering algorithm needs to be performed to group these subpaths into a number of clusters. The clustering of subpaths depends on their delays, powers, and AoAs. In principle, AoDs can be considered as well. However, in this paper, we considered clustering from a receiver's perspective. Therefore, the cluster algorithm is not related to AoDs. The K-means++ algorithm \cite{Arthur07} was developed based on the K-means algorithm with an additional simple randomized seeding technique. Here, the K-means++ algorithm was employed to analyze the number of clusters.

Euclidean distance was used as the metric for clustering. Let $K$ be the total number of centroids. The K-mean++ clustering algorithm first randomly chooses $K$ centroids. Then, it group a subpath into a cluster by the minimum Euclidean distance, i.e., 
\begin{align}
\hat{k}_j=\underset{k\in\left\lbrace 1,2,\cdots,K \right\rbrace}{\mathrm{argmin}}\sqrt{\beta^2(\tau_j-\tau_k)^2+(\theta^\mathrm{A}_j-\theta^\mathrm{A}_k)^2+(\theta^\mathrm{E}_j-\theta^\mathrm{E}_k)^2}
\end{align}  
where $\beta=10$ is a scaling coefficient, $\tau_k$, $\theta^\mathrm{A}_k$, and $\theta^\mathrm{E}_k$ are the delay, azimuth AoA, and elevation AoA of the $k$th centroid, $\tau_j$, $\theta^\mathrm{A}_j$, and $\theta^\mathrm{E}_j$ are the delay, azimuth AoA, and elevation AoA of the $j$th subpath, and $\hat{k}_j$ is the estimated cluster index of the $j$th subpath. Then, each centroid is updated by averaging subpaths inside and the K-means++ clustering algorithm continues iteratively until convergence.
Finally, the optimum number of clusters $N=K_{\mathrm{opt}}$ is determined by the Kim-Park (KP) index in \cite{Kim01}. The KP index is computed by the mean intra-cluster distance and the inter-cluster minimum distance. The introduction of KP index is beyond the scope of this paper, details can be found in \cite{Kim01}.
\subsection{Number of subpaths per cluster}
After clustering the subpaths, the number of subpaths $M_n$ within the $n$th cluster can be obtained. Let $W_n=\{j:\hat{k}_j=n\}$, then $M_n=|W_n|$. It can be observed that the numbers of subpaths within clusters are not necessarily the same.

\subsection{Intra-cluster delay spread}
In \cite{Salous13}, the definition of delay spread for clusters was given. In this paper, this definition can be generalized to compute intra-cluster delay spreads. The average intra-cluster delay $\bar{\tau}_n$ and root mean square (RMS) delay spread $S_n$ for the $n$th cluster can be expressed as
\begin{align}
\bar{\tau}_n=\frac{\sum\limits_{k=1}^{M_n}\tau_{nk}P_{nk}}{\sum\limits_{k=1}^{M_n}P_{nk}}-\tilde{\tau}_n
\end{align} \label{equ_avg_delay}
\begin{align}
S_n=\sqrt{\frac{\sum\limits_{k=1}^{M_n}\left(\tau_{nk}-\bar{\tau}_n-\tilde{\tau}_n\right)^2P_{nk}}{\sum\limits_{k=1}^{M_n}P_{nk}}}
\end{align} \label{equ_delay_spread}
where $\tilde{\tau}_n=\min\left\lbrace \tau_{n1},\tau_{n2},... \right\rbrace$ and $P_{nk}$ is the subpath power. Since the intra-cluster RMS delay spreads are not equal for different clusters, cumulative distribution function (CDFs) are needed to study their statistical properties.

\subsection{Intra-cluster angular spread}
Similar to intra-cluster delay spreads, the average intra-cluster azimuth AoA $\bar{\theta}^{\mathrm{A}}_n$ and azimuth AoA spread $Q_n$ for the $n$th cluster can be expressed as
\begin{align}
\bar{\theta}^{\mathrm{A}}_n=\frac{\sum\limits_{k=1}^{M_n}\theta^{\mathrm{A}}_{nk}P_{nk}}{\sum\limits_{k=1}^{M_n}P_{nk}}
\end{align} \label{equ_avg_AAoA}
\begin{align}
Q_n=\sqrt{\frac{\sum\limits_{k=1}^{M_n}\left(\theta^{\mathrm{A}}_{nk}-\bar{\theta}^{\mathrm{A}}_n\right)^2P_{nk}}{\sum\limits_{k=1}^{M_n}P_{nk}}}.
\end{align} \label{equ_delay_spread}
Let $V_n$ be the intra-cluster elevation AoA spread of the $n$th cluster. The same procedure can be used to compute $V_n$. However, it is neglected here to avoid repeated interpretations.

\section{Results and Analysis} \label{sec_results_and_analysis}

The numbers of clusters of the 28 GHz NLOS channel in street canyon are listed in Table \ref{tab_summary}. The mean number for ray tracing simulation is 2.2 while the mean number for measurement is 3.4. The reason for that ray tracing has a smaller number of clusters than measurement is because ray tracing simulations were performed in a simplified environment. Hence, certain strong scattering objects such as lampposts were ignored. Another important observation is that the total number of clusters at 28 GHz is significantly less than that at below 6 GHz channels. Similar conclusions can be found in measurement performed by New York university \cite{Akdeniz14}. The default value of the number of clusters in NLOS is $20$ in the WINNER II channel model \cite{winner}. It should be reduced for above 6 GHz channel models. Based on the simulation and measurement results, a constant number of three clusters would be a reasonable starting point to balance complexity and accuracy.

\begin{figure}[t]
\centering
\includegraphics[width=3.5in]{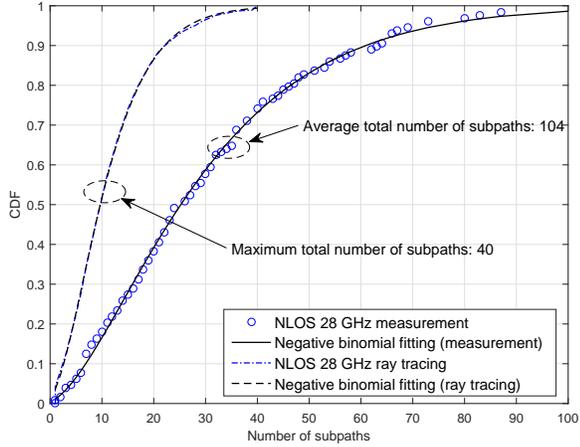}    
\caption{CDFs of number of subpaths for both ray tracing and measurement.}
\label{fig_Nsubpath_fitting}
\end{figure}

CDFs of numbers of subpaths within one cluster are depicted in Fig. \ref{fig_Nsubpath_fitting}. The mean value of ray tracing is $11.8$, which is significantly less than the mean value $31.0$ of measurement. One reason for this is that the maximum total number of subpaths considered in ray tracing simulations is limited by $40$. If the maximum total number of subpaths is scaled to $104$, same as the average total number of subpaths in measurements, the scaled mean number of subpaths within one cluster would be $11.8\times\frac{104}{40}=30.7$. This is aligned with measurement results. In the current WINNER II channel model, the number of subpaths within a cluster is constant $20$, which may not be feasible for above 6 GHz channel models. Also, it can be seen in Fig. \ref{fig_Nsubpath_fitting} that the number of subpaths within one cluster follows a negative binomial distribution, which is different from the uniform distribution assumption used in \cite{Samimi15}. Then, the probability mass function (PMF) $f(M_n)$ of the number of subpaths within one cluster can be presented as
\begin{align}
f(M_n)=\frac{\Gamma(M_n+r)}{M_n!\Gamma(r)}\cdot p^{M_n}(1-p)^{r}
\end{align} \label{equ_avg_delay}
where $r$ is positive real number and $0<p<1$.

\begin{figure}[t]
\centering
\includegraphics[width=3.5in]{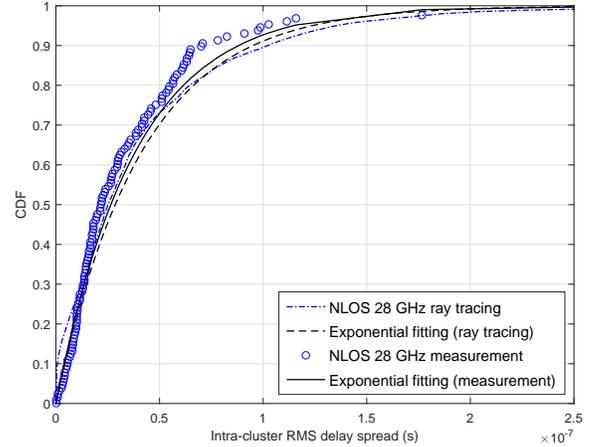}    
\caption{CDFs of intra-cluster RMS delay spreads for both ray tracing and measurement.}
\label{fig_IntraCluster_delay_spread_fitting}
\end{figure}

Next, CDFs of intra-cluster RMS delay spreads are illustrated in Fig.  \ref{fig_IntraCluster_delay_spread_fitting}. Both ray tracing and measurement results match well. The mean and median values of the intra-cluster RMS delay spread are approximately $40$ ns and $22$ ns. These are larger than the current assumption in the WINNER II channel model, which is $10$ ns. Therefore, channel models for above 6 GHz channels should take intra-cluster delay spreads into account. Exponential distributions can be applied to fitting the distribution of intra-cluster delay spreads.

\begin{figure}[t]
\centering
\includegraphics[width=3.5in]{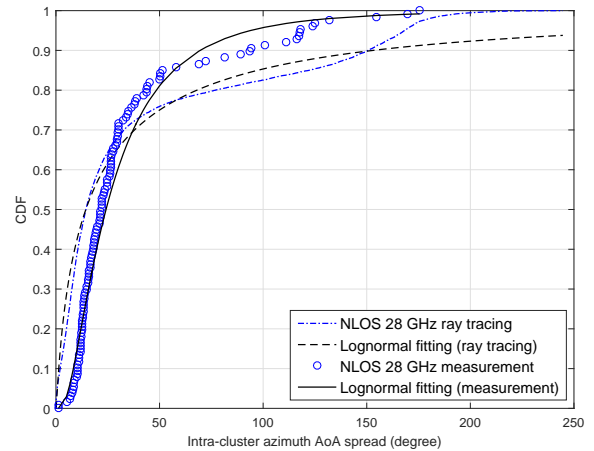}    
\caption{CDFs of intra-cluster azimuth AoA spreads for both ray tracing and measurement.}
\label{fig_AAoA_spread_28GHz_CDF_MeaVsRaytracing_fitting}
\end{figure}

\begin{figure}[t]
\centering
\includegraphics[width=3.5in]{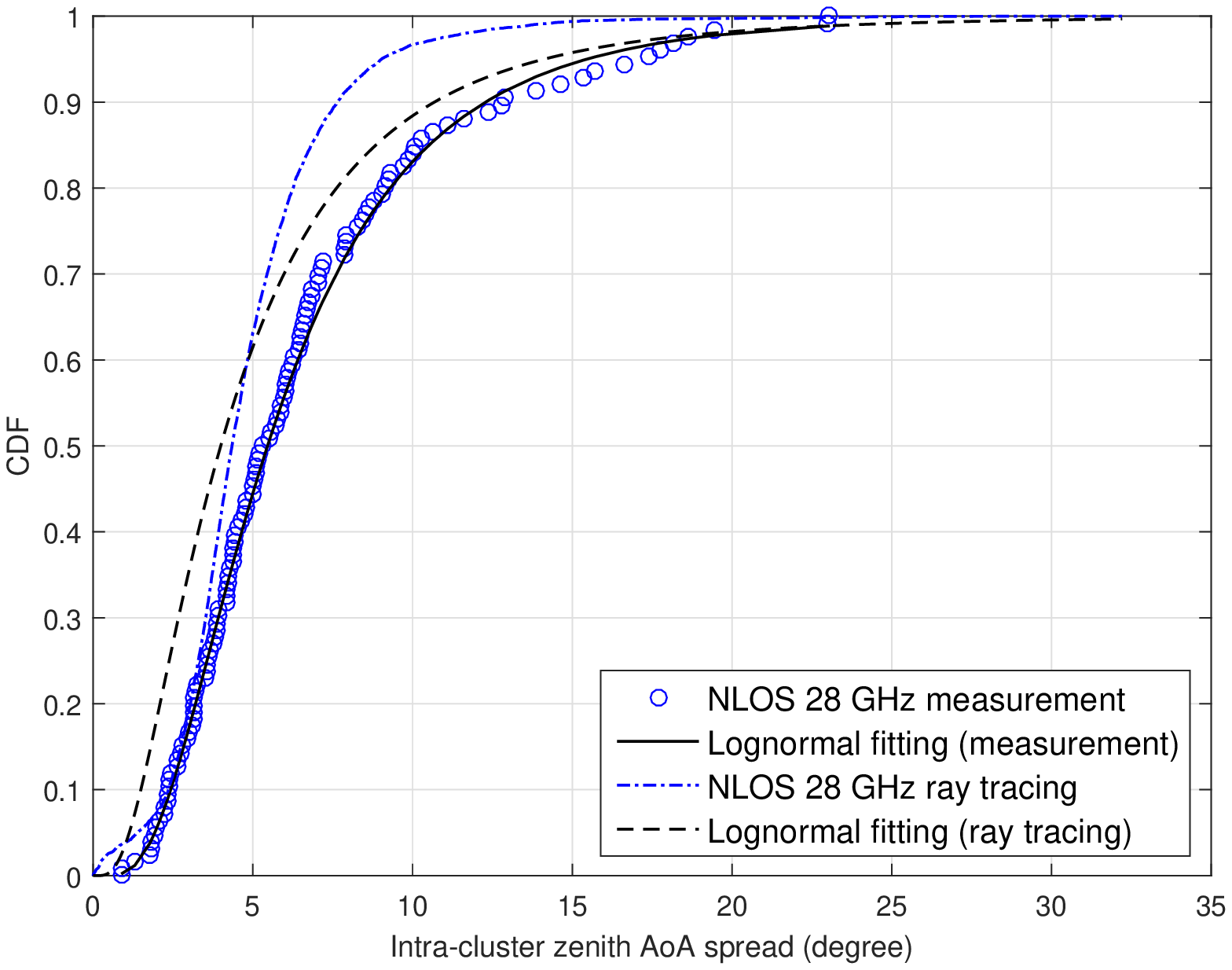}    
\caption{CDFs of intra-cluster zenith AoA spreads for both ray tracing and measurement.}
\label{fig_ZAoAspread_fitting}
\end{figure}

In Fig. \ref{fig_AAoA_spread_28GHz_CDF_MeaVsRaytracing_fitting} and Fig. \ref{fig_ZAoAspread_fitting}, CDFs of intra-cluster angular spreads are presented. Since 3-D effects are considered, angular spreads should consist of azimuth angular spreads and elevation angular spreads. In Fig. \ref{fig_AAoA_spread_28GHz_CDF_MeaVsRaytracing_fitting}, ray traced azimuth AoA spread and measured azimuth AoA spread have a similar trend, their mean values are approximately $40$ degrees. Also, similar trend can be observed in intra-cluster zenith AoA spread in Fig. \ref{fig_ZAoAspread_fitting}. However, it can be observed that there is a small gap between the ray tracing result and the measurement result in Fig. \ref{fig_ZAoAspread_fitting}. This can be caused by that zenith angles are not fully stochastic and they have certain dependency on the distance between the transmitter and receiver. Locations of ray tracing and measurement do not fully overlap, which is responsible for the gap. However, the gap is within $2^{\circ}$ which is sufficiently small to be neglected. Also, lognormal distributions can be used to model angular spreads in both azimuth and elevation.
 
Summary of proposed models including probability density functions (PDFs) or PMFs and parameters for intra-cluster characteristics is listed in Table \ref{tab_summary}.

\begin{table*}[t]
\caption{Summary of models and parameters for intra-cluster characteristics.}
\center
\normalsize
    \begin{tabular}{|c|c|c|c|c|}
    \hline
  &      \multicolumn{4}{|c|}{Ray tracing}  \\ \hline
  & Proposed model (PMF/PDF)   & Mean & Median & Parameters \\ \hline
        No. of clusters & Constant & 2.2 & 2 & -  \\ \hline
        No. of subpaths per cluster & $f(M_n)=\frac{\Gamma(M_n+r)}{M_n!\Gamma(r)}\cdot p^{M_n}(1-p)^{r}$ & $11.8$ & $10$ & $p=0.18$, $r=2.63$  \\ \hline
        Intra-cluster delay spread &$f(S_n)=\lambda e^{-\lambda S_n}$ & $41.3$ns & $25.1$ns & $\lambda=2.42\times 10^7$   \\ \hline
 Intra-cluster azimuth AoA spread &$f(Q_n)=\frac{1}{Q_n\sigma\sqrt{2\pi}}\exp\left(-\frac{(\ln Q_n -\mu)^2}{2\sigma^2} \right)$  & $41.4^\circ$ & $14.5^\circ$ & $\mu=2.67$, $\sigma=1.84$   \\ \hline
    Intra-cluster zenith AoA spread & $f(V_n)=\frac{1}{V_n\sigma\sqrt{2\pi}}\exp\left(-\frac{(\ln V_n -\mu)^2}{2\sigma^2} \right)$ & $4.77^\circ$ & $4.34^\circ$ & $\mu=1.38$, $\sigma=0.77$  \\ \hline
  &       \multicolumn{4}{|c|}{Measurement} \\ \hline
        No. of clusters & Constant & 3.4 & 3 & - \\ \hline
        No. of subpaths per cluster & $f(M_n)=\frac{\Gamma(M_n+r)}{M_n!\Gamma(r)}\cdot p^{M_n}(1-p)^{r}$ & $31.0$ & $26$ & $p=0.06$, $r=1.96$   \\ \hline
       Intra-cluster delay spread & $f(S_n)=\lambda e^{-\lambda S_n}$ & $38.2$ns & $21.9$ns & $\lambda=2.62\times 10^7$   \\ \hline
Intra-cluster azimuth AoA spread & $f(Q_n)=\frac{1}{Q_n\sigma\sqrt{2\pi}}\exp\left(-\frac{(\ln Q_n -\mu)^2}{2\sigma^2} \right)$  & $34.9^\circ$ & $22.1^\circ$ & $\mu=3.18$, $\sigma=0.82$   \\ \hline
       Intra-cluster zenith AoA spread & $f(V_n)=\frac{1}{V_n\sigma\sqrt{2\pi}}\exp\left(-\frac{(\ln V_n -\mu)^2}{2\sigma^2} \right)$  & $6.66^\circ$ & $5.40^\circ$ & $\mu=1.70$, $\sigma=0.63$  \\ \hline
 \hline
    \end{tabular}
    \label{tab_summary}
\end{table*}

\section{Conclusions} \label{conclusion_section}
Intra-cluster characteristics for a NLOS street canyon scenario at 28 GHz have been investigated in this paper. The analysis is based on data obtained via both ray tracing simulations and measurements. Results have shown that the number of clusters is smaller than that in the below 6 GHz WINNER II channel model. Also, certain intra-cluster characteristics such as the number of subpaths, intra-cluster delay spreads and angular spreads are required to be included for future WINNER based above 6 GHz channel models. Distributions of these intra-cluster characteristics have been provided. Moreover, it has been shown that ray tracing simulations are able to provide similar results to measurements. Therefore, ray tracing simulations can be used as reference when channel measurements are not available. In future work, channel models at 28 GHz based on the revealed intra-cluster characteristics are to be developed. Furthermore, it will be interesting to see intra-cluster characteristics in LOS conditions.

\section*{Acknowledgment}
\small
The research leading to these results received funding from the European Commission H2020 programme under grant agreements $\mathrm{n}^{\circ}$671650 (5G PPP mmMAGIC project).

\end{document}